\documentclass[conference]{IEEEtran}

\DeclareSymbolFont{bbold}{U}{bbold}{m}{n}
\DeclareSymbolFontAlphabet{\mathbbold}{bbold}
\makeatletter
\newcommand\semihuge{\@setfontsize\semihuge{22.5}{22.9}}
\makeatother
\usepackage{algpseudocode}
\usepackage{algorithm}
\usepackage{algorithmicx}

\usepackage{lipsum} % For algorithm smape after or before:

\usepackage[dvips]{color}
\usepackage{comment}
\usepackage{todonotes}
\usepackage{epsf}
\usepackage{epsfig}
\usepackage{times}
\usepackage{epsfig}
\usepackage{graphicx}
\usepackage{bbold}
\usepackage{mathtools}
\usepackage{mathrsfs}
\usepackage{amssymb}
\usepackage{pdfpages}
\usepackage{epstopdf}
\usepackage{float}
\newfloat{algorithm}{t}{lop}
\usepackage{amsmath}

\usepackage{dsfont}
\usepackage{lettrine} % \lettrine[findent=1pt]{{{R}}}{}
\usepackage{amsmath,epsfig,amssymb,algorithm,algpseudocode,amsthm,cite,url}
\usepackage{subcaption}
\allowdisplaybreaks
\usepackage{csquotes}
\usepackage{verbatim}
\usepackage[english]{babel}
\usepackage{amsmath,amssymb}

\usepackage[justification=centering]{caption}
\usepackage{verbatim}

\IEEEoverridecommandlockouts

\begin{document}
	
\title{Environment-Aware Deployment of Wireless Drones Base Stations with Google Earth Simulator\vspace{-0.01cm}}    

\author{\IEEEauthorblockN{\large Aaron French, Mohammad Mozaffari, Abdelrahman Eldosouky, and Walid Saad}\vspace{-0.25cm}\\
	\IEEEauthorblockA{
		 Wireless@VT, Electrical and Computer Engineering Department, Virginia Tech, VA, USA, \\Emails:\url{{ajfrench, mmozaff, iv727, walids}@vt.edu}.
	%	\thanks{This research was supported by the U.S. National Science Foundation under Grants CNS-1460316 and CNS-1638283.}
		%\thanks{ This work was supported in part by the Army Research Office (ARO) under Grant 	W911NF-17-1-0593, in part by the US NSF under Grant AST-1506297, by the ERC Starting Grant MORE (Advanced Mathematical Tools for Complex Network Engineering), and by Academy of Finland (CARMA).}
	}\vspace{-0.60cm}}
\maketitle%\vspace{-0.5cm}
%\vspace{-0.4cm}
\begin{abstract}\vspace{-0.02cm}	
	In this paper, a software-based simulator for the deployment of base station-equipped unmanned aerial vehicles (UAVs) in a cellular network is proposed. To this end, the Google Earth Engine platform and its included image processing functions are used to collect geospatial data and to identify obstacles that can disrupt the line-of-sight (LoS) communications between UAVs and ground users. Given such geographical information, three environment-aware optimal UAV deployment scenarios are investigated using the developed simulator. In the first scenario, the positions of UAVs are optimized such that  the number of ground users covered by UAVs is maximized. In the second scenario, the minimum number of UAVs needed to provide full coverage for all ground users is determined. Finally, given the load requirements of the ground users, the total flight time (i.e., energy) that the UAVs need to completely serve the ground users is minimized. Simulation results using a real area of the Virginia Tech campus show that the proposed environment-aware drone deployment framework with Google Earth input significantly enhances the network performance in terms of coverage and energy consumption, compared to  classical deployment approaches that do not exploit geographical information. In  particular, the results show that the proposed approach yields a coverage enhancement by a factor of 2, and a 65\% improvement in energy-efficiency. The results have also shown the existence of an optimal number of drones that leads to a maximum wireless coverage performance.    %Essentially, this paper functions as a case study on the viability of Google Earth Engine as a platform assisting planners of UAV-enabled networks. We determine that ...
\end{abstract} \vspace{0.2cm}

\section{Introduction}

Unmanned aerial vehicles (UAVs), or drones have recently attracted significant attention as a promising approach to enhance wireless communication performance \cite{mozaffari2018tutorial, ALZ1,WuConference,bor,mozaffari2018beyond}. When equipped with wireless base station hardware, drones can supplement the coverage provided by existing cellular infrastructure. The mobility of drones facilitates the creation of line-of-sight (LoS) links with users, ensuring optimal connection strength. This ability, coupled with the reliability and autonomy of drones, lends UAVs attractive qualities to service providers. In particular, UAVs are an effective approach in emergency scenarios such as disaster relief, when unplanned power outages may compound with the increased need for communication, and Internet of Things (IoT) applications \cite{mozaffari2}, where the quantity and low transmit power of devices may necessitate closer-ranged wireless communications. Meanwhile, UAVs can also be used to complement existing terrestrial cellular systems by bringing additional capacity to crowded areas during temporary events. Furthermore, drones can be deployed to provide necessary wireless connectivity to rural areas in which the presence of large-scale ground wireless infrastructure is limited.

To effectively deploy drones drone base stations in wireless networking applications, there is a need for efficient simulators that can simulate different use-case scenarios and ground environments. Though many simulators have been developed for terrestrial base stations \cite{I1}, \cite{I2}, only some are suited specifically for the analysis of three-dimensional, ad hoc networks  \cite{I3}. These are typically implemented as extensions of the general network simulators \cite{I4}, that operate in two dimensions. UAV-enabled networks are highly dynamic and thus require a proper integration of the movement of UAVs into the simulation environment. Moreover, analysis of these networks is made more challenging by the uncertainty of environmental variables affecting propagation, as well as highly dynamic interference. To account for these UAV features, many models implement probabilistic expressions based on environment type, i.e., rural, urban, or dense urban \cite{Letter}. Thus, the ability to identify obstacles by processing satellite images can have
  immense value in that drone simulations can become more deterministic, depending on the accuracy of image processing.

While there has been a notable number of works on UAV
deployment, they do not consider the potential use of real
geographical information for optimal placement of UAVs. For instance, the work in \cite{HouraniOptimal} optimizes the altitude of a single UAV for maximizing coverage based on a probabilistic path loss model. Using this model, the authors in \cite{Kalantari} studied the coverage maximization problem with minimum number of drone base stations. In \cite{Azari2}, the deployment of  an aerial UAV base station for maximizing sum-rate and power gain in a wireless network is studied. These studies use variations of the probabilistic models introduced above, and thus, are not suited for simulation of real-world environments. In contrast to previous studies on UAV deployment,  we  extract environmental information with great precision by using image processing tools in Google Earth Engine. Subsequently, we build a drone deployment  simulator that accepts buildings' locations as inputs, and adaptively determines the optimal positions of the drones for maximizing wireless connectivity in various scenarios.  %Unlike the prior studies on the deployment of drones, we will exploit geographical information such as obstacles to effectively deploy and operate drone-enabled wireless systems.    

The main contribution of this paper is a novel simulation framework for \emph{environment-aware} deployment of multiple drone base stations that provide wireless connectivity for ground users. In particular, by exploiting geographical information extracted from the Google Earth Engine, we determine the locations of buildings that disrupt LoS. Then, we use our simulator to investigate three key UAV deployment scenarios. First, we study the optimal placement of drones for maximizing the number of covered ground users. In the second scenario, we aim to provide full coverage for ground users by using a minimum number of drones. Finally, given the load requirements of users, we analyze the optimal deployment of drones for which the total flight time of drones needed to service the users is minimized. Simulation results reveal that our proposed framework  that exploit buildings' information on an area in Virginia Tech's campus using Google Earth  yields a significant improvement in the coverage and energy efficiency of the drone-enabled wireless networks. Moreover, our results show the existence of an optimal number of drones that maximizes the wireless connectivity.

The rest of this paper is organized as follows. In Section II, we present the system model the drone deployment scenarios. In Section III, we describe the developed feature (i.e., obstacle) extraction method from Google Earth. Simulation results are presented in Section IV and conclusions are drawn in Section V. \vspace{-0.00cm}

\section{System Model and Drone Deployment Scenarios }

\begin{figure}[!t]
	\begin{center}
		\includegraphics[width=8.7cm]{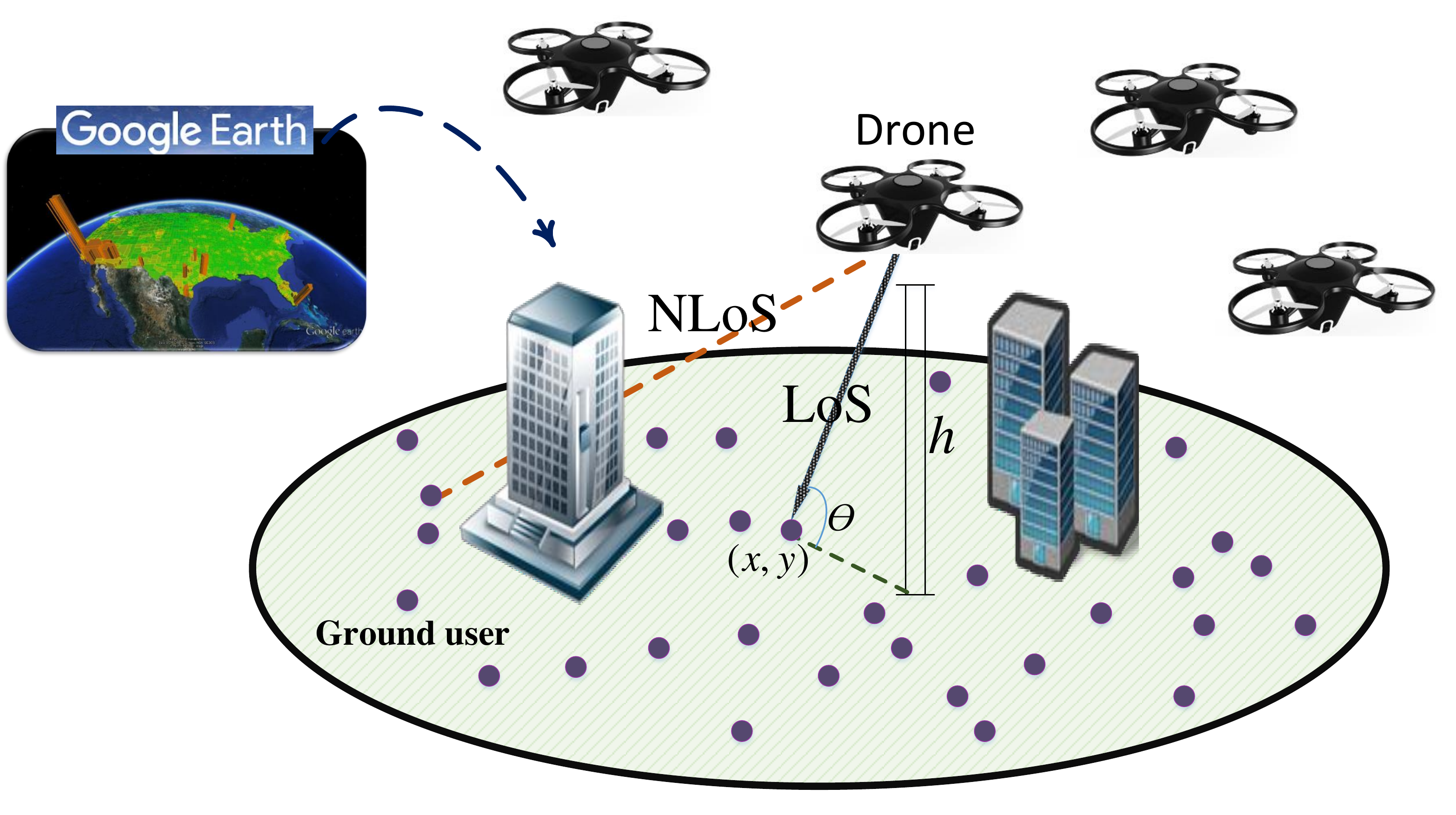}
		\vspace{-0.1cm}
		\caption{  System model for drones' deployment.} \vspace{-.4cm}
		\label{SystemModel}
	\end{center}	
\end{figure}

Consider a set $\mathcal{L}$ of $L$ single-antenna wireless users located within a given geographical area. The location of a user $i \in \mathcal{L}$ is given by $(x_i,y_i)$. In this area, a set $\mathcal{M}$ of $M$ quadrotor drones are used as flying base stations to provide downlink wireless service to ground users, as shown in Figure \,\ref{SystemModel}. The location of a drone $j \in \mathcal{M}$ is given by $\boldsymbol{v}_j = (x_j^D,y_j^D,h_j)$. 

Each user $i$ can be served by one drone $j$ that provides the strongest downlink signal-to-interference-plus-noise-ratio (SINR) for the user such that $\gamma_{ij} = \mathop {\arg \max }\limits_{j \in \mathcal{M}} \gamma_{ij}$ and $\gamma_{ij} \ge \gamma_{\textrm{th}}$ where $\gamma_{ij}$ is the SINR downlink between user $i$ and drone $j$ and $\gamma_{\textrm{th}}$ is threshold SINR required by the user to successfully have wireless service. Here, the SINR for user $i$ that connects to drone $j$ can be given by:
\begin{align}
&\gamma_{ij}=\frac{\eta P_j  d_{ij}^{-\alpha}}{{\sum\limits_{u \in {\mathcal{I}_\textrm{int}}} {\eta {P_u}d_{u}^{-\alpha}}}+ \sigma^2},\\
&d_{ij}=\sqrt{(x_i-x_j^D)^2+(y_i-y_j^D)^2+h_j^2},
\end{align}
where $\alpha$ is the path loss exponent, $\sigma^2$ is the noise power, $\eta$ is the path loss constant. $d_{ij}$ is the distance between drone-BS $j$ and a given user $i$. Also, $\mathcal{I}_\textrm{int}$ is the set of interfering drone-BSs.

We assume that users have fixed locations and that drones can move to certain locations to service the users. Our goal is to optimally deploy the drones, i.e., calculate optimal locations to provide the wireless service in each of the following scenarios.

\subsection{Maximizing the Number of Covered Users}

In the first scenario, our goal is to maximize the number of covered users under limited resources (available drones). This scenario captures emergency scenarios, e.g., flooding or power outage, or highly unusual wireless service demand, e.g., a fair or a sports event in a stadium. In such cases, the goal of using drones is to provide wireless service to the largest possible number of users. Covering every user in these cases might not be possible due to very high data demand that will require more drones than what is available. Determining the number of drones that can be used in these scenarios depends on the number of available drones and the expected coverage in this geographical area. In emergency cases for example, when more than one geographical area is affected and in need for urgent coverage, drones are to be deployed in these areas according to the percentage of ground users that can be effectively covered with connectivity by the drones.

In this scenario, the number of users is fixed to $L$ and the number of drones is fixed to $M$. The goal is to find the optimal locations of the drones $\boldsymbol{v}_j, \  \forall j \in \mathcal{M}$ to maximize the number of covered users. Let $\mathbbold{1}_{ij}$ be an indicator of whether or not user $i$ is connected to drone $j$ such that:\vspace{0.1cm}
\begin{equation}
\mathbbold{1}_{ij} = 
\begin{cases}
%\begin{array}{lr}
1 \hspace{0.5cm}  \textrm{if }  j = \mathop {\arg \max }\limits_{j \in \mathcal{M}} \gamma_{ij}$ and $\gamma_{ij} \ge \gamma_{\textrm{th}},\\
0 \hspace{0.5cm}  \textrm{if } \textrm{otherwise.}
%  \end{array}
\end{cases}
\end{equation}

The problem can then be formulated as:
\begin{align}
\mathop {\max}_{\mathcal{L}} &\sum_{i \in \mathcal{L}} \sum_{j \in \mathcal{M}} \mathbbold{1}_{ij} \\
\label{eq:constraint1_1} \text{s. t.}  \hspace{0.5cm}& \sum_{j \in \mathcal{M}} \mathbbold{1}_{ij} = 1, \forall i \in \mathcal{L}.
\end{align}

The constraint in (\ref{eq:constraint1_1}) guarantees that every user is connected to only one drone.

\subsection{Full Coverage with a Minimum Number of Drones}

In this next scenario, every user needs to be covered using the minimum number of drones. Here, unlike the previous scenario, we do not assume limited resources. This scenario usually occurs in public safety and pre-disaster awareness situations in which every user needs to be informed by a disaster mitigation plan.  For example, in pre-disaster evacuation, we need to make sure that every user is aware of the upcoming danger in a timely-manner. This can help improve the community resilience against these type of disasters.  Covering every user (i.e., full coverage) can be achieved by deploying drones in the targeted geographical area. However, as deploying these drones is usually costly, we need to ensure full coverage while minimizing the number of drones, and, hence the cost.

The goal is to calculate the minimum number of drones required to achieve full user coverage to the $L$ available users. This is achieved by calculating the optimal locations of the drones $\boldsymbol{v}_j, \  \forall j \in \mathcal{M}$ to achieve full coverage of the users. We use the same indicator $\mathbbold{1}_{ij}$ as defined in the previous scenario. The problem can then be formulated as: \vspace{0.2cm}
\begin{align}
\mathop {\min}_{\mathcal{M}} & \sum_{j \in \mathcal{M}} \sum_{i \in \mathcal{L}} \mathbbold{1}_{ij} \\
\text{s. t.}  \hspace{0.5cm}& \sum_{j \in \mathcal{M}} \mathbbold{1}_{ij} = 1, \forall i \in \mathcal{L},  \\
& \sum_{i \in \mathcal{L}} \mathbbold{1}_{ij} = L.
\end{align}

The first constraint ensures that every user is connected to only one drone and the second constraint ensures that all the users are connected to drones.

\subsection{Minimizing Flight Time of Drones in Serving Users}

In this third scenario, each user needs to download some data using the wireless service and we are interested in minimizing the hover time (service time) of the drones to satisfy this data load for every user. This scenario captures the case in which the consumed energy is of importance as the drones can provide wireless services for only a limited period of time \cite{MozaffariFlightTime}. One example scenario is the case in which the drones are to be deployed in a geographical area that is far from their source and, thus, the drones will have to consume a significant portion of their energy for traveling to the destination. The remaining amount of energy (that will be used to serve the users) needs to be used in the most effective way possible so as to satisfy the demand of the users.

Each user, among the $L$ users, is assumed to have a load of data given by $\beta_i$ bits that needs to be satisfied. A drone $j$ can transmit data to a user $i$ with a rate $b_{ij}$ bits/second that depends on $\gamma_{ij}$.  The time spent by a drone $j$ to serve a user $i$ can then be calculated as:
\begin{equation}
t_{ij} = \frac{\beta_i}{b_{ij}}.
\end{equation}

The total hover time of a drone $j$ can then be calculated as the summation of the times spent to serve all the users connected to this drone. Let $\mathcal{N}_j$ be the set of all users connected to drone $j,  \ \forall j \in \mathcal{M}$. Then, the hover time for a drone $j \in \mathcal{M}$ will be is given by:
\begin{equation}
t_j =    \sum_{i \in \mathcal{N}_j} \frac{\beta_i}{b_{ij}}.
\end{equation}
The goal in this third scenario is to find the optimal locations of the drones to minimize the overall hover time of all drones given that the load of each user needs to be satisfied. The problem can be formulated as:
\begin{align}\label{eq:scenario3}
\mathop {\min}_{\mathcal{M}} &\sum_{j \in \mathcal{M}} \sum_{i \in \mathcal{N}_j} \frac{\beta_i}{b_{ij}} \\
\text{s. t.}  \hspace{0.5cm}& \sum_{j \in \mathcal{M}} \mathbbold{1}_{ij} = 1, \forall i \in \mathcal{L}  \\
& \sum_{i \in \mathcal{L}} \mathbbold{1}_{ij} = L.
\end{align}

The constraints are similar to the previous scenario. In this scenario, when every user is connected to a drone, then every user will be in a set $\mathcal{N}_j$ of a specific drone $j$ such that:
\begin{equation}
\bigcup_{j \in \mathcal{M}} \mathcal{N}_j = \mathcal{L}.
\end{equation}

Then, the problem formulation of~(\ref{eq:scenario3}) will minimize the overall hover time while ensuring that the total load of users is satisfied. \vspace{0.3cm}

\section{Google Earth Engine Simulator for Obstacle Location Extraction}

To analyze the aforementioned scenarios, using a ground environment-aware approach, we have developed a drone network simulator using MATLAB that takes as input the locations of buildings. To determine these, we now explore the use of Google Earth Engine, a platform that is suitable for analysis and representation of geospatial data. Earth Engine incorporates multiple datasets and image processing tools. The simplest way to use the Earth Engine is through its built-in JavaScript IDE, which we explore in this work. Python is also supported through an API. The platform is well-suited for our application because of the image processing potential, allowing us to estimate and refine network parameters, and the intuitive interface through which users can supervise the building detection process.

Various building detection algorithms have been developed, with cited precisions ranging from 80-90\% \cite{E1,E2,cohen2016rapid}. Accurate algorithms rely on a combination of feature extraction techniques and machine learning. For our application, we circumvent the time and resources needed to train such programs by exploiting the “map view” imagery supplied by Google. In this view, satellite imagery is simplified, wherein features like buildings are identified in the same color. This greatly facilitates automated building identification, under the assumption that Google's own identification techniques are accurate.

To extract building locations from map view, we use edge detection. This is implemented most readily in Earth Engine through Canny edge detection, a reliable and very common algorithm \cite{E6,duda1972use}. Canny detection applies separate filters to detect horizontal, vertical, and diagonal edges, and computes the gradient magnitude. Finally, non-maximum magnitudes are suppressed, thinning the detected edges. In general applications of edge detection, image noise must be accounted for through the application of Gaussian filters; even then, error is expected. However, the simple, noiseless images provided by map view are ideal candidates for edge detection, and edge detection yields accurate results.

\begin{figure}[!t]
	\begin{center}
		\includegraphics[width=8.5cm]{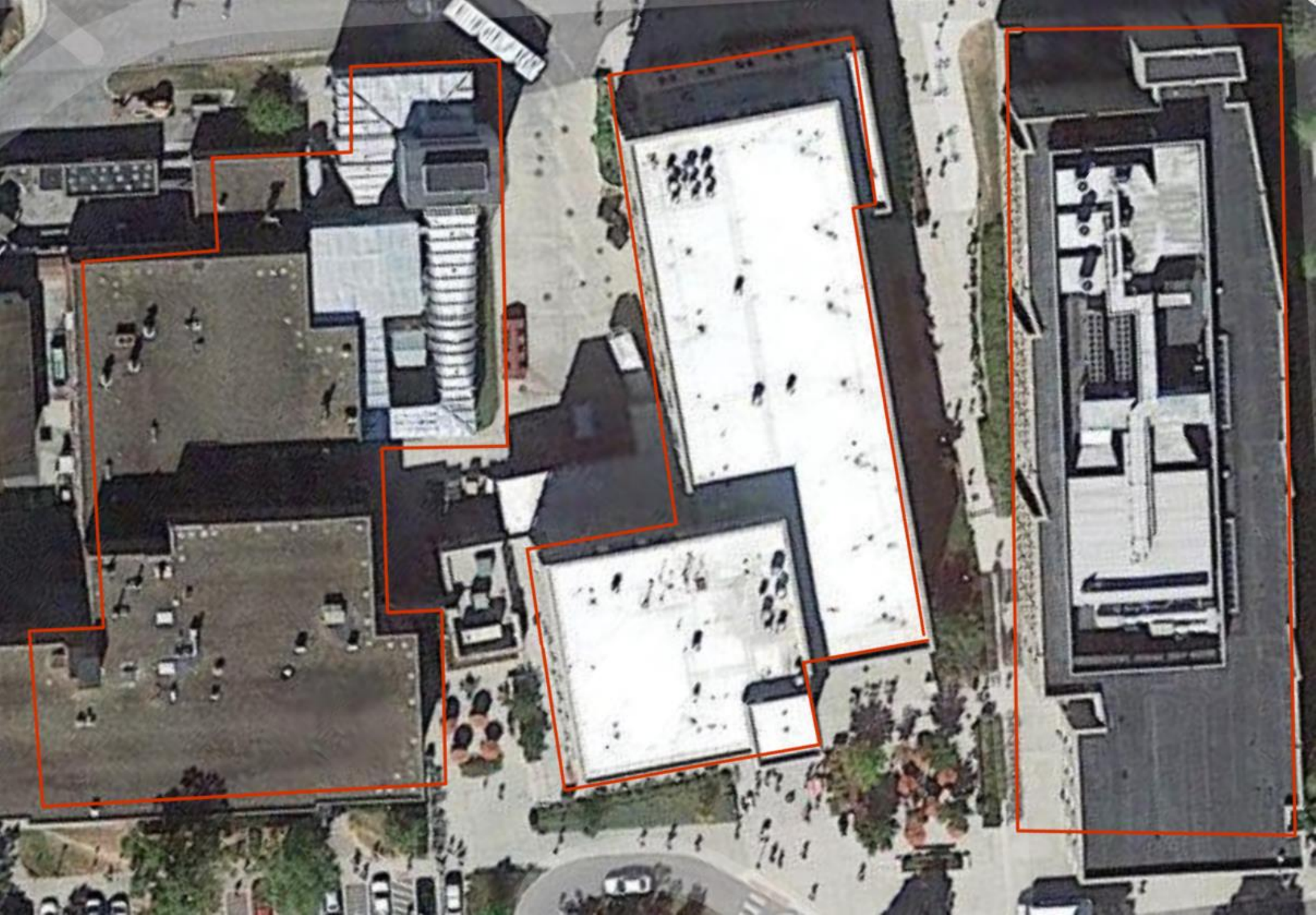}
		\vspace{0.02cm}
		\caption{ Results of building identification imposed over satellite imagery for a region of the Virginia Tech campus.} \vspace{-.6cm}
		\label{Image_Google}
	\end{center}	
\end{figure}

To extract lines from this output, we apply the Hough transform to the Canny image \cite{E6}. This step is important to correct imperfections in the Canny output. The Hough transform uses an accumulator to detect the presence of a line, then implements a voting algorithm to identify its parameters. Now, we sample and trace each line, noting changes in direction which correspond to building corners. At this point, we can also manually adjust the locations of any vertex. To examine the accuracy of this process we outlined buildings on map view and overlaid them onto the corresponding satellite imagery, shown in Figure\,\ref{Image_Google}.

Evidently, through this method, buildings are approximated fairly well. Over five test cases that we performed, this process correctly outlined about 95\% of each building's correct area, and falsely identified an additional 12\%. These figures are consistent with the 80-90\% accuracy bounds given in the studies cited above. Additionally, we note that this method tends to overestimate building area. This is permissible, and possibly preferable, for UAV simulations, in which drones should not be deployed within a buffer area around buildings, due to the threat of collision. The limitations of the geometric approximation of buildings in this manner include irregular building shapes, specifically ones with rounded sides. Earth Engine only supports polygons; thus, rounded edges must be represented by some number of vertices, adding inherent error. In summary, we have shown that for building location identification, analysis of Google map data is consistent in accuracy with rigorous processing of satellite imagery, but can be performed at reduced computational cost. Thus, while using minimal computational resources, we have identified the locations of buildings, which will be used as inputs into our developed Google Earth-enabled MATLAB simulator so as to analyze the proposed environment-aware wireless drone base station deployment scenarios. \vspace{0.05cm}

%To extract building locations from map view, we use edge detection. Geospatial analysis in Earth Engine is performed on images, fundamental data structures representing raster data. So, we first create an image from the area of interest. Then, after loading the image, we use the Hough transform in conjunction with Canny edge detection to extract edges of buildings \cite{E6}. Finally, we approximate each building as a polygon, and find the coordinates of each vertex. At this point, we can also manually adjust the vertices of the shapes, further improving accuracy. To examine the accuracy of this process we outlined buildings on map view and overlaid them onto the corresponding satellite imagery, shown in Figure\,\ref{Image_Google}.

%Evidently, through this method, buildings are approximated fairly well. This was true over various test cases; moreover, we found that the outputted shapes tend to overestimate building area. This is permissible, and possibly preferable, for UAV simulation, in which drones should not be deployed within a buffer area around buildings, due to the threat of collision. Limitations of geometric approximation of buildings include irregular building shapes, specifically ones with rounded sides. Earth Engine only supports polygons; thus, rounded edges must be represented by some number of vertices, adding inherent error.

\section{Simulation Results}
For our simulations, we consider a 200\,m $\times$ 200\,m area over which users are randomly distributed. Users are assumed to be at ground level, at which $z = 0$. The locations of buildings are known, defined by their vertices at $\{V_1, V_2,...,V_N\}$, where each vertex consists of an $x$ and $y$ coordinates. For these simulations, we consider a three-building configuration derived from Figure 2 which is based on a real area from the Virginia Tech campus. As we did not estimate building height during image processing, we model the buildings' $z$-coordinates as random variables, constrained between 10 and 20 meters, heights appropriate for five-story buildings. Other simulation parameters are listed in Table 1.

To evaluate any arrangement of $M$ drones over $N_C$ candidate points, ${N_C}\choose{M}$ calculations are required. As $N_C$ correlates directly with simulation precision, and hence a large $N_C$ is desirable, the computational complexity can quickly become infeasible. To circumvent this, the following heuristic is implemented. We first discretize the target area into some number $N_C$ of UAV candidate points, where $N_C$ is sufficiently small to enable rapid evaluation. We form the binary power threshold matrix $\boldsymbol{T}$ in which entry $(m,n)$ indicates whether the user at location $(x_n, y_n)$ receives above a given power $P^t_\textrm{min}$ from candidate point $m$. Note that we do not yet account for interference, noise, or line-of-sight; our current goal is to establish starting points for further optimization. We incrementally place drones at the candidate points is maximized; in other words, at points with the most potential links.

Now, we further discretize the area around each chosen candidate point. Given $\{V_1, V_2,...,V_N\}$, we calculate whether a LoS exists by sampling the line segment connecting user $i$ and each candidate point and checking whether any sample point lies within the bounds of a building. If so, we introduce an additional attenuation factor, $\eta$, to that potential channel. Finally, we consider interference and noise, and simultaneously solve for the optimal locations of each UAV such that the number of users above a given SINR threshold, $\gamma$, is maximized.

\begin{figure}[!t]
	\begin{center}
		\includegraphics[width=9.2cm]{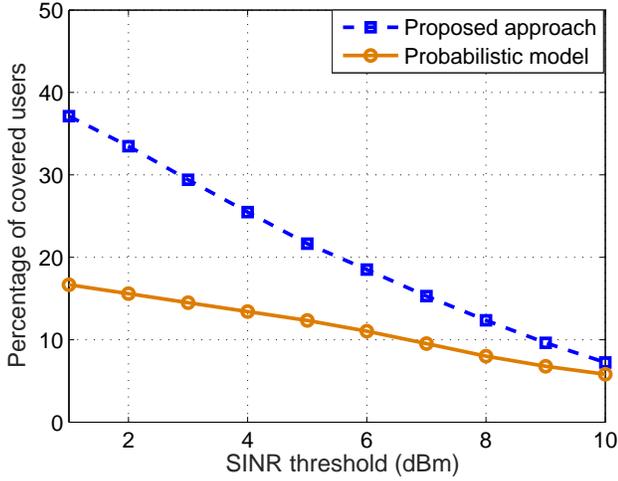}
		\vspace{-0.3cm}
		\caption{ Percentage of covered users versus SINR threshold.} \vspace{-.3cm}
		\label{SINRvsUsers}
	\end{center}	
\end{figure}

Figure \ref{SINRvsUsers} shows the percentage of covered users as the SINR threshold needed for connectivity varies (this result corresponds to the deployment scenario in Subsection II-$A$). Clearly, as the SINR threshold or equivalently the receivers' sensitivity increases, the coverage performance of drones decreases. This due to the fact that satisfying a higher SINR requirement is more challenging thus fewer number of users can be covered by the drones. For instance, when increasing the SINR threshold from 2\,dBm to  8\,dBm, the number of covered users decreases by 63\% in the proposed approach. In Figure \ref{SINRvsUsers}, we also compare the performance of the proposed deployment approach with a case in which deployment is done based a probabilistic path loss model. In the probabilistic model, a drone can have a LoS link to a ground user with a specific probability, which is given by \cite{HouraniModeling}:
\begin{equation} \label{PLoS}
{P_{\text{LoS},i}} = b_1 {\left( {\frac{180}{\pi}\theta_i  - 15} \right)^{b_2} }, %\,\,\, \theta_i>\frac{\pi}{12},\vspace{-0.1cm}
\end{equation}
where $\theta_i$ is the elevation angle (in radians) between the drone $i$ and a user located at $(x,y)$. Also, $b_1$ and $b_2$ are constant values which depend on the environment.

\begin{figure}[!t]
	\begin{center}
		\includegraphics[width=8.3cm]{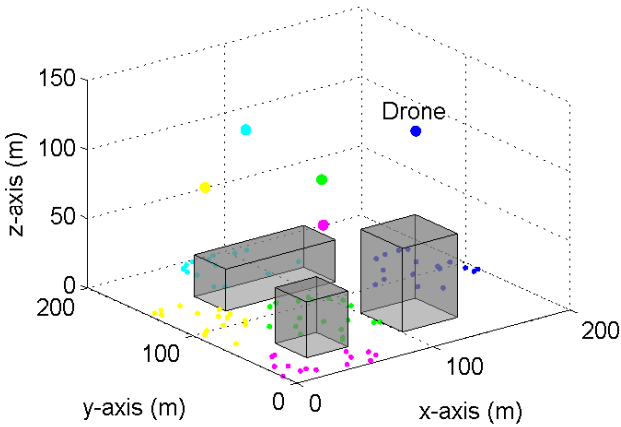}
		\vspace{-0.2cm}
		\caption{ An illustrative figure for drones' deployment.} \vspace{-.4cm}
		\label{UAV_graphic_det}
	\end{center}	
\end{figure}

As we can see from Figure \ref{SINRvsUsers}, our approach outperforms the probabilistic case. In our approach, the locations of buildings are known and deterministic as they are obtained from the Google Earth Engine. In the probabilistic case, however, we do not have a complete information about the buildings. Therefore, by exploiting additional information  about the environment, our deployment approach leads to a higher coverage performance than the probabilistic-based deployment. As shown in Figure \ref{SINRvsUsers}, the number of covered users can be increased by up to a factor of 2 while adopting the proposed environment-aware deployment strategy. As an illustrative example, in Figure \ref{UAV_graphic_det}, we show visual output of drone placement, using known building locations.            

\begin{figure}[!t]
	\begin{center}
		\includegraphics[width=9.3cm]{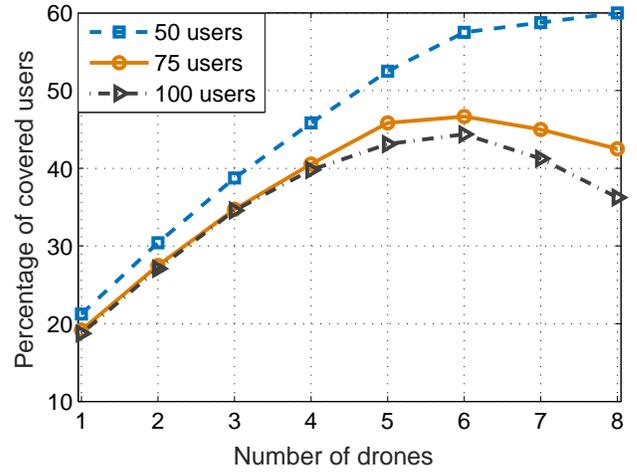}
		\vspace{-0.4cm}
		\caption{Percentage of covered users versus number of drones.} \vspace{-.5cm}
		\label{dronesvsUsers}
	\end{center}	
\end{figure}

Figure \ref{dronesvsUsers} shows the impact of the number of drones on the coverage performance for various network sizes (this result corresponds to the deployment scenario in Subsection II-$B$). Clearly, the coverage performance decreases as the number ground users increases. For a higher number of users, it will be more likely that drone-users communication links will become blocked by obstacles. Consequently, the communication reliability and, hence, the coverage performance degrades. Figure \ref{dronesvsUsers} also shows how the number of covered users varies by changing the number of drones. In this case, there is a tradeoff in deploying more drone base stations for providing wireless connectivity. By increasing the number of drones, the coverage can be improved as the drones are placed closer the ground users. However, while using more drones, the aggregated interference increases which reduces the users' SINR. Therefore, there exists an optimal number of drones for which the coverage is maximized. For instance, as we can see from Figure \ref{dronesvsUsers}, the optimal number of drones for serving 100 users is 6. This figure allow us to determine the minimum number of drones needed to meet a certain coverage requirement. For example, here, a full coverage for 50 users can be achieved by optimally deploying 8 drones over the considered geographical area.

\begin{figure}[!t]
	\begin{center}
		\includegraphics[width=9.2cm]{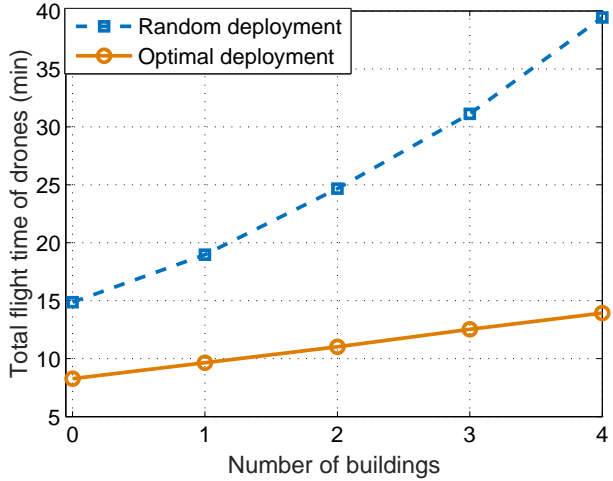}
		\vspace{-0.3cm}
		\caption{Total flight time of drones versus number of buildings (i.e., obstacles).} \vspace{-.1cm}
		\label{flightTime}
	\end{center}	
\end{figure}

Figure \ref{flightTime} shows the total flight time of drones needed for completely servicing the users (this result corresponds to the deployment scenario in Subsection II-$C$). From this figure, we can see that the flight time of drones increases when the number of buildings (i.e., obstacles) increases. With more obstacles in the environment, drone-to-user communications will experience lower SINR due to the blockage and shadowing effects. As a result, the transmission rate will decrease and the drones must fly longer in order to transmit a required amount of  data to each user. From Figure \ref{flightTime}, it can be seen that the total flight time of drones increases by 45\%, in the proposed deployment case, when the number of buildings increases from 1 to 4. Hence, servicing users located in a harsh environment requires longer flight time, more energy consumption, and thus using more capable drones. 

In Figure \ref{flightTime}, we compare the performance of our proposed environment-aware deployment approach with a random deployment case in which drones are randomly deployed over the geographical area. As we can see from Figure \ref{flightTime}, the proposed optimal deployment can yield up to a 65\% flight time reduction compared to the random deployment case. Therefore, the proposed approach enhances energy-efficiency of the considered drone-enabled wireless network.  \vspace{0.1cm}

\begin{table}[!t]
	\normalsize
	\begin{center}
		%\centering
		\caption{ Simulation parameters.}
		\vspace{-0.1cm}
		\label{TableI}
		\resizebox{8.8cm}{!}{
			\begin{tabular}{|c|c|c|}
				\hline
				\textbf{Parameter} & \textbf{Description} & \textbf{Value} \\ \hline \hline
				$f_c$	&     Carrier frequency     &      2\,GHz     \\ \hline 
				$P_i$	&     Drone transmit power     &     1\,\textrm{W}    \\ \hline
				
				$N_o$	&     Total noise power spectral density     &        -170\,dBm/Hz  \\ \hline
				%	$\gamma$	&     SINR threshold      &   5\,dB \\ \hline
				$N$	&     Number of ground users     &   200 \\ \hline

				$B$	&    Bandwidth      &   1\,MHz \\ \hline
				
				$b_1,b_2$	&    Parameters in probabilistic channel model & $0.36, 0.21$ \cite{HouraniOptimal}\\ \hline
				
				$\beta$	&    Load per ground user      &   10\,Mb \\ \hline
			\end{tabular}}
			
		\end{center}\vspace{-0.60cm}
	\end{table}

\section{Conclusion}
In this paper, we have investigated the problem of environment-aware deployment of drone base stations that provide wireless connectivity to ground users. To this end, first, we have developed a drone network simulator that uses the
 Google Earth Engine in order to extract key information about buildings in the considered geographical area. Then, we have studied the optimal deployment of drones in three practical scenarios. In the first scenario, we have determined the optimal locations of drones such that the number of covered ground users is maximized. In the second scenario, we have minimized the number of drones needed to ensure a full coverage for all users. Finally, we have minimized the flight time of drones required to completely service the users by satisfying their load requirements. Our results have shown that the proposed deployment framework significantly enhances the drone wireless system performance in terms of coverage and energy efficiency. Moreover, our simulation results have demonstrated existence of an optimal number of drones for which the wireless coverage is maximized.   \vspace{0.00cm}   
%In this paper, we have proposed a novel framework for employing a drone-enabled antenna array system that can provide high rate (i.e., low service time) wireless services to ground users. In particular, we have optimized the positions of drones (as the array elements) within the antenna array such that the service time for each user is minimized. Our results have shown that the proposed drone antenna array with the optimal configuration yields a significant improvement in terms of the service time, spectral and energy efficiency.\vspace{-0.01cm} 
%\end{comment}
 
%\newpage
\def\baselinestretch{1.01}
\bibliographystyle{IEEEtran}

\bibliography{references}

\vspace{-0.02cm}
% that's all folks
\end{document}